# Mechanical properties of PMMA-sepiolite nanocellular materials with a bimodal cellular structure

Victoria Bernardo[1]*, Frederik Van Loock[2], Judith Martin-de Leon[1], Norman A. Fleck[2] and Miguel Angel Rodriguez-Perez[1]

1. Cellular Materials Laboratory (CellMat), Condensed Matter Physics Department, University of Valladolid, Valladolid, Spain

2. Engineering Department, University of Cambridge, Trumpington Street, CB2 1PZ Cambridge, United Kingdom

*Corresponding author: Victoria Bernardo (vbernardo@fmc.uva.es) +34 983 18 40 35

**ABSTRACT**

Nanocellular PMMA with up to 5 wt% of nano-sized sepiolites are produced by gas dissolution foaming. The porosity of 50% to 75% exists in a bimodal cell size distribution with micro- and nano-sized cells. Uniaxial compression tests are performed to measure the effect of sepiolite concentration on the elastic modulus and the yield strength of the solid and cellular nanocomposites. Single edge notch bend tests are conducted to relate the fracture toughness of the solid and cellular nanocomposites to sepiolite concentration too. The relative modulus is found to be independent of sepiolite content to within material scatter when considering the complete porosity range. In contrast, a mild enhancement of the relative modulus was observed by the addition of sepiolite particles for the foamed nanocomposites with a porosity close to 50%. The relative compressive strength of the cellular nanocomposites is found to mildly decrease as a function of sepiolite concentration. A strong enhancement of the relative fracture toughness by the addition of sepiolites is observed. The enhancement of the relative fracture toughness and the relative modulus (at 50% porosity) can be attributed to an improved dispersion of the particles due to foaming and the migration of micro-sized aggregates from the solid phase to the microcellular pores during foaming.

1. **INTRODUCTION**

Nanocellular polymers are polymer foams characterised by cell sizes in the range of tens to hundreds of nanometres. An attractive property of these nanocellular polymers is their low thermal conductivity due to the Knudsen effect [1,2]. Recently, semi-transparent nanocellular foams have been reported [3,4] and, due to their nano-sized cell size, these materials have the potential to be used in membranes for ultrafiltration or in catalysis and sensors [5–7]. Most research on nanocellular polymers is focused on their production, whereas the literature on the mechanical characterisation is relatively scarce. Notario et al. [8] found that the material performance index for a light, stiff beam in bending $E^{1/2}/\rho$ (where E is the Young's modulus and ρ is the density) for a nanocellular foam exceeded that for a microcellular foam [9]. They attributed this stiffening to the fact that the size of the cell walls of the nanocellular material is in the order of the radius of gyration of a PMMA molecule [8]. Miller and co-workers [10] found that micro- and nanocellular polyetherimide (PEI) have similar values for E whereas the nanocellular PEI materials had a greater impact resistance. Guo [11] observed that micro- and nanocellular polycarbonate (PC) have similar values of E/ρ and similar impact resistance properties for cellular materials with relative densities higher than 0.6.

The addition of inorganic nanoparticles to a polymer matrix is a common strategy to improve the mechanical properties of a polymer [12–15]. When these nanocomposites are foamed, the resulting cellular nanocomposites inherit this reinforcement and this strategy could be used to further enhance the mechanical properties of nanocellular foams [16,17]. In addition, nano-sized particles have successfully been used as heterogeneous nucleation agents for the production of micro- [18–21] and nanocellular [22–26] polymers. The addition of nanoparticles is therefore a promising method to enhance the mechanical performance of nanocellular polymers. However, the authors have been unable to locate any studies that investigate the effect of nanoparticles on the mechanical properties of nanocellular polymers.

In the present study, nanocellular polymethylmethacrylate (PMMA) is reinforced with nano-sized needle-like sepiolites. The effect of the sepiolite concentration on the mechanical properties (such as the compressive yield strength, the compressive elastic modulus, and the fracture toughness) of the solid and cellular nanocomposites is measured. In an earlier work [26] we showed that the addition of sepiolites, modified with a quaternary ammonium salt, in a PMMA matrix resulted in bimodal cellular structures comprizing micro- and nano-sized cells. In this paper, our goal is to analyse the mechanical behaviour of these bimodal nanocellular polymers and to determine the effect of the addition of sepiolites particles on their mechanical properties.

## 2. EXPERIMENTAL

### 2.1. Materials

Polymethylmethacrylate (PMMA) V 825T ($M_n$ = 43 kg/mol, $M_w$ = 83 kg/mol) was supplied by ALTUGLAS® International in the form of pellets with a density (ρ) of 1.18 g/cm$^3$ and a glass transition temperature ($T_g$) close to 114.5 °C as measured by DSC. Sepiolites were provided by Tolsa S.A (Spain). These particles are hydrated magnesium silicates. Sepiolites present a needle-like morphology, with an average particle length ranging from 1 μm to 2 μm and a diameter in the nanometre range (between 20 nm and 30 nm) [27,28]. The sepiolites used in this work have been modified with a quaternary ammonium salt. The process to obtain and modify these particles is detailed elsewhere [29,30]. Medical grade carbon dioxide ($CO_2$) (99.9% purity) was used as the blowing agent for the gas dissolution foaming experiments.

### 2.2. Solid blends production

Blends of PMMA with varying sepiolite contents were compounded using a twin-screw extruder model COLLIN TEACH-LINE ZK 25T, with L/D equal to 24 and screw diameter equal to 25 mm (**Table 1**). PMMA and sepiolites were dried in a vacuum oven at 50 °C for 12 hours before blending. The temperature profile set on the extruder was from 160 °C at the hopper to 200 °C in the die. The screw speed was equal to 40 rpm. The produced blends were cooled in a water bath and pelletized. After drying the pellets for 2 hours in a vacuum oven at a temperature equal to 50 °C, each blend was extruded again using the same processing conditions to have a homogeneous dispersion of the particles.

Next, the obtained pellets were compression moulded into solid sheets of 155x75x4 mm$^3$ using a hot plate press provided by Remtex. The pellets were first dried in a vacuum oven at 50 °C overnight before processing. Subsequently, they were made molten by holding them at 250 °C for 500 s and then compacted at 250 °C with a constant pressure of 17 MPa for 60 s. Finally, the sheets were cooled down to room temperature with the pressure of 17 MPa maintained. Rectangular specimens with dimensions corresponding to 50x15x4 mm$^3$ were machined from the sheet for the foaming experiments. Note that PMMA absent the sepiolite was processed under the same conditions for comparison purposes.

Table 1. The PMMA-sepiolite blend formulations.

| Material ID | Sepiolite concentration (wt%) |
|---|---|
| PMMA | 0 |
| 1%-S | 1 |
| 2%-S | 2 |
| 3%-S | 3 |
| 5%-S | 5 |

### 2.3. Gas Dissolution Foaming Experiments

Foaming experiments were performed using a pressure vessel (model PARR 4681) provided by Parr Instrument Company with a capacity of 1 litre. The maximum temperature and pressure reached by the pressure vessels correspond to 350 °C and 41 MPa, respectively. The pressure is automatically controlled by a pressure pump controller (model SFT-10) provided by Supercritical Fluid Technologies Inc. The vessel is equipped with a clamp heater of 1200 W, and its temperature is regulated via a CAL 3300 temperature controller. Foaming experiments were conducted by a two-step foaming process [31]. First, samples were put into the pressure vessel at a constant $CO_2$ pressure ($p_{sat}$ = 10 MPa) and temperature ($T_{sat}$ = 25 °C) for the saturation stage. At these conditions, full saturation of $CO_2$ in PMMA is achieved within 20 hours [26]. The pressure was progressively released to ambient pressure with a controlled pressure drop rate of 15 MPa/s.

The foaming step was carried out in a hot and cold plates press from Remtex [32]. Details about this foaming process can be found in the **Supplementary Information**. To obtain materials with different densities, the temperature of the press and the foaming time were varied (see **Table 2**). After the foaming step in the hot and cold plates press, flat samples, suitable for mechanical characterization, were obtained. From these pieces, samples with adequate dimensions for the different mechanical tests were machined. For the blend with the highest particle content (5%-S), it was only possible to produce the materials with high relative densities, as the presence of too many aggregates of the sepiolites particles led to cracking of the samples at the highest foaming temperatures used to produce the low and medium relative density cellular nanocomposites.

Table 2. Foaming parameters in the press.

| Target relative density | Temperature (°C) | Time (s) |
|---|---|---|
| High (~ 0.5) | 40 | 300 |
| Medium (~ 0.35) | 60 | 300 |
| Low (~ 0.3) | 100 | 60 |

## 2.4. Characterization
### 2.4.1. Density

The density of the solid nanocomposites was measured with a gas pycnometer (Mod. AccuPyc II 1340, Micromeritics). The density of the cellular materials was determined with the water-displacement method based on the Archimedes' principle using a density determination kit for an AT261 Mettler-Toledo balance. The solid skin of the samples was removed with a polisher (model LaboPOl2-LaboForce3, Struers) by polishing off 200 μm from the top and bottom faces of the sample before measuring their densities. The relative density ($\rho_r$) is defined as the ratio of the cellular material density ($\rho$) to the density of the solid nanocomposite with the same composition ($\rho_s$).

### 2.4.2. Cellular Structure

Samples were immersed in liquid nitrogen and then fractured for microscopic visualization and coated with gold using a sputter coater (model SCD 005, Balzers Union). The cellular structure of the samples was analysed using an ESEM Scanning Electron Microscope (QUANTA 200 FEG). Dedicated in-house software based on ImageJ/FIJI was used for this purpose [33]. Firstly the average cell size ($\phi$) was measured and the standard deviation of the cell size distribution ($SD$) was obtained. The parameter $SD/\phi$ was calculated as an indicator of the homogeneity of the cellular structure. The nanocomposite cellular materials of this work possess a bimodal cellular structure with micro-sized cells (above 1 μm) and nano-sized cells (below 1 μm), and values for the average cell size $\phi$ and standard deviation $SD$ were measured for both distributions. We write the average cell size as $\phi_1$ for the nano-sized cells and as $\phi_2$ for the micro-sized cells. Similarly, $SD_1$ refers to the standard deviation of the cell size distribution of the nano-sized cells and $SD_2$ denotes the standard deviation of the cell size distribution of the micro-sized cells. The anisotropy ratio $AR$ was measured as the ratio between the average cell size of the whole population of cells observed in the plane aligned with the compression moulding direction to the average cell size of the whole population of cells measured in the plane perpendicular to the compression moulding direction. Cell density ($N_v$) and cell nucleation density ($N_0$) were determined from the SEM images using Kumar's theoretical approximation [34] according to:

$$N_v = \left[\frac{n}{A}\right]^{3/2} \quad (1)$$

$$N_0 = \frac{N_v}{\rho_r} \qquad (2)$$

where $n$ is the number of cells in the SEM image and $A$ is the area of the image. Note that more than 200 cells from various regions of each cellular material were analysed.

In this work, bimodal cellular structures (with cell sizes in the micro and the nano scale) are obtained. The observed cellular structures were found to have a much larger proportion of nano-sized cells than micro-sized cells. The micro-sized cells, however, typically occupied a significant volume of the sample, in the range from 20% to 40%. To quantify the observed bimodality, the relative volume occupied by the population of nano-sized cells, $V_{nano}$, is measured [35]:

$$V_{nano} = \frac{A_t - A_m}{A_t} \qquad (3)$$

where $A_m$ is the observed area occupied by the micro-sized cells (cell size above 1 μm) in the SEM images, and $A_t$ the total area of the image. The resulting two-dimensional area ratio should be representative for the three-dimensional volume ratio when an adequate amount of surfaces are analysed, according to stereology [36,37].

### 2.4.3. Open Cell content

The open cell content of the cellular materials was measured according to the ASTM D6226-10 standard using a gas pycnometer (Mod. AccuPyc II 1340, Micromeritics). The open cell content ratio $OC$ is defined as:

$$OC = \frac{V - V_p - V_s}{V(1 - \rho_r)} \qquad (4)$$

where $V$ is the geometric volume of the sample, $V_p$ is the volume measured by the pycnometer and $V_s$ is a penalty volume to account for the exposed cells at the surface of the sample. The geometric volume was determined from the cellular material density (measured by the water-displacement method) and its mass ($m$) (measured with an AT261 Mettler-Toledo balance) as $V = m/\rho$. $V_p$ was determined by performing a pressure scan (from 0.02 MPa to 0.13 MPa) in the gas pycnometer and measuring the pycnometric volume for each pressure. It was assumed that no more gas is able to enter the interconnected open cells when the measured volume remains constant for an increase in pressure. $V_p$ was calculated as the average of these last measured constant volume values. Note that, as $V_s$ is proportional to the cell size, this value becomes negligible for micro and nanocellular materials.

### 2.4.4. X-Ray analysis

X-ray imaging is employed to determine the number of particle aggregates in the nanocomposite material. For this purpose, X-ray tomography images of both solid and cellular materials were taken with a spatial resolution of 2.5 μm (i.e. aggregates with dimensions larger than 2.5 µm can be detected) [38]. The mass of the aggregates is calculated by measuring the volume occupied by the aggregates and taking into account the real volume fraction of particles in the sample.

In addition, all samples were analysed by X-ray radiography [39] and those samples presenting defects or inhomogeneities were excluded from the mechanical tests.

### 2.4.5. Mechanical tests

Mechanical properties in uniaxial compression were measured using an Instron 5584 electromechanical testing machine. Specimens were cuboids with in-plane dimensions 10x10 mm$^2$; the thickness varied from 4 mm to 6 mm depending on the relative density of the sample. The compression direction was perpendicular to the compression moulding direction. At least three specimens were tested per material system. Tests were carried out at a crosshead velocity equal to 0.5 mm min$^{-1}$, corresponding to a strain rate equal to 8.3 x 10$^{-4}$ s$^{-1}$. Displacement of the platens was measured via a laser extensometer. All tests were conducted at room temperature.

Single edge notch three point bending (SENB) tests were performed at room temperature with an Instron 5584 test bench at a constant crosshead speed of 10 mm/min. Specimens were cuboids with in-plane dimensions 55x15 mm$^2$; the thickness varied from 4 mm to 6 mm depending on the relative density of the sample. The critical mode I stress intensity factor $K_{Ic}$ was calculated as a measure for the fracture toughness in accordance with the ASTM D5045-14 [40]. A pre-crack with a sharp tip was made at the end of a sawed notch by tapping with a razor blade.

## 3. RESULTS AND DISCUSSION

### 3.1. Cellular structure

Representative cellular structures of the cellular materials with a relative density close to 0.5 are shown in **Figure 1**. At very low magnification (first row of **Figure 1**), a homogeneous structure is observed for PMMA, whereas the blends with sepiolites have a heterogeneous structure with pore sizes exceeding 100 µm. Using a higher magnification, one can observe the microcellular structure of the pure PMMA and the blends with sepiolites (see second row of **Figure 1**). The cell size distribution of the pure PMMA is unimodal; there are no nano-sized pores present (see third row of **Figure 1** where an even higher magnification is used**)**. In contrast, the PMMA/sepiolite blends have a bimodal cell size distribution, the dominant population of cells is nanocellular as detailed below (see the third row in **Figure 1**). Earlier work demonstrated that sepiolites modified with a quaternary ammonium salt act as a nucleating agent during gas dissolution foaming of PMMA [26]. It was suggested that the microcellular pores appear due to micro-sized sepiolite aggregates. The well-dispersed sepiolites account for the presence of nanocellular pores.

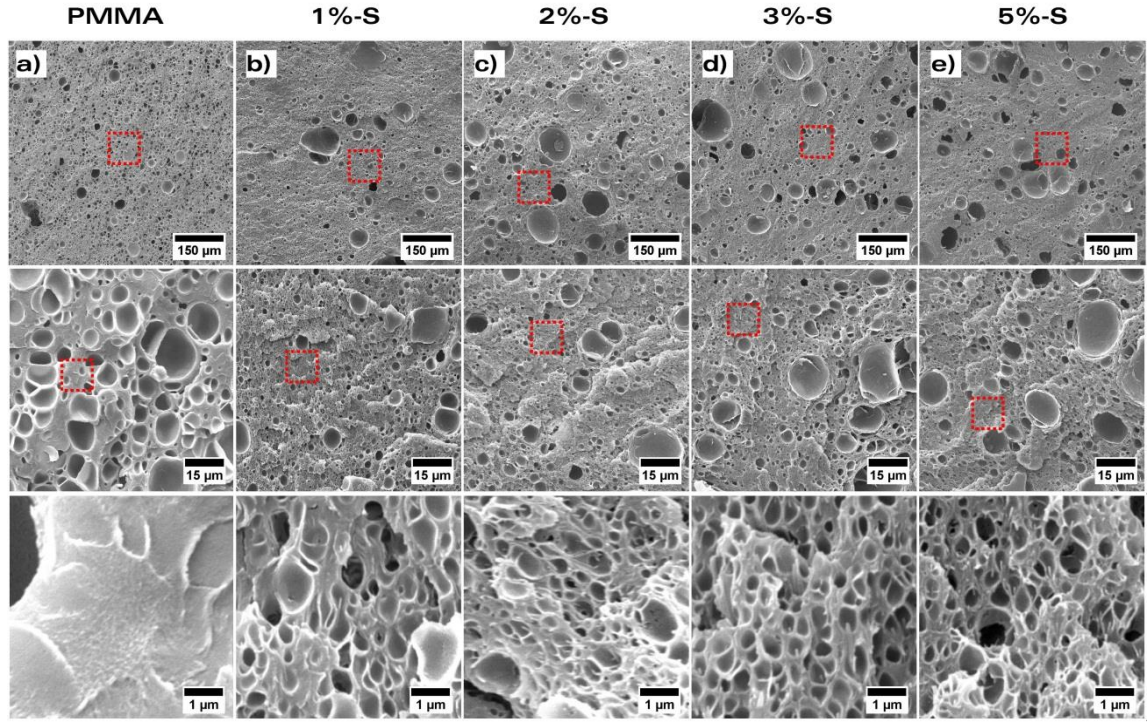

**Figure 1.** SEM images of the samples produced at a foaming temperature equal to 40 ºC and foaming time equal to 5 minutes: a) PMMA, b) 1%-S, c) 2%-S, d) 3%-S and e) 5%-S. The second and third rows show images of the same materials at increased magnification.

The main parameters characterizing the cellular structure of all the material systems produced in this study are summarized in **Table 3**. Due to the difference in size between the largest and the smallest cells in the materials with bimodal cell size distribution, we identify two sets of cells: the main (nanocellular) and the secondary (microcellular) structures. The microcellular pores were measured using SEM micrographs with the magnification of the images shown in the second row of **Figure 1** (cell size around 1-10 µm). The volumetric fraction of nano-sized cells ($V_{nano}$ in **Table 3**) is greater than 50% for all the materials and, for this reason, the nanocellular population is considered to be the dominant one.

Bimodal micro- and nanocellular materials with average cell sizes ranging from 330 nm to 500 nm in the nano-sized cell population are obtained, whereas for the micro-sized cell population the cell size ranges from 3 µm to 7 µm. The nanocellular cell populations are more homogeneous, with $SD_1/\phi_1$ values around 0.5-0.7, while the microcellular population is strongly heterogeneous with values for $SD_2/\phi_2$ higher than 1. It is observed that, for the high-density materials (samples 1 to 5), an increased sepiolite content leads to a mild reduction of the average cell size. For the lower density materials, this effect is less obvious. For a given sepiolite concentration, the cell size tends to increase when density is reduced. Regarding the cell nucleation density, an increase of the nucleation in three orders of magnitude with respect to the pure PMMA is detected when sepiolites are added. The cellular materials were found to be closed-celled as the measured open cell contents were lower than 10% for all the material systems. In addition, the materials can be considered as isotropic because the anisotropy ratio is close to 1 for all the systems under study.

**Table 3.** Measured cellular structure parameters and open cell content of the cellular samples produced in this work.

| # | Material | Relative Density | Cell Nucleation Density (nuclei/cm³) | $V_{nano}$ | $\phi_1$ (nm) | $SD_1/\phi_1$ | $\phi_2$ (μm) | $SD_2/\phi_2$ | $AR$ | $OC$ |
|---|---|---|---|---|---|---|---|---|---|---|
| 1 | PMMA | 0.52 ± 0.04 | $2.12 \cdot 10^{10}$ | 0.00 | 4268 | 0.77 | - | - | 1.1 ± 0.4 | 0.077 |
| 2 | 1%-S | 0.50 ± 0.02 | $1.15 \cdot 10^{13}$ | 0.75 | 456 | 0.51 | 3.4 | 0.92 | 1.3 ± 0.5 | 0.097 |
| 3 | 2%-S | 0.51 ± 0.02 | $2.88 \cdot 10^{13}$ | 0.79 | 345 | 0.52 | 3.7 | 1.08 | 1.0 ± 0.5 | 0.057 |
| 4 | 3%-S | 0.53 ± 0.01 | $2.00 \cdot 10^{13}$ | 0.61 | 332 | 0.72 | 3.1 | 1.07 | 1.1 ± 0.5 | 0.086 |
| 5 | 5%-S | 0.47 ± 0.02 | $3.11 \cdot 10^{13}$ | 0.66 | 307 | 0.66 | 3.6 | 1.02 | 1.1 ± 0.4 | 0.059 |
| 6 | PMMA | 0.35 ± 0.01 | $6.95 \cdot 10^{10}$ | 0.0 | 3209 | 0.92 | - | - | 1.2 ± 0.5 | 0.065 |
| 7 | 1%-S | 0.38 ± 0.04 | $1.61 \cdot 10^{13}$ | 0.71 | 436 | 0.56 | 4.0 | 0.74 | 1.1 ± 0.4 | 0.041 |
| 8 | 2%-S | 0.35 ± 0.01 | $1.16 \cdot 10^{13}$ | 0.55 | 422 | 0.69 | 5.0 | 1.00 | 1.2 ± 0.5 | 0.049 |
| 9 | 3%-S | 0.35 ± 0.01 | $1.33 \cdot 10^{13}$ | 0.61 | 419 | 0.72 | 7.2 | 0.80 | 1.4 ± 0.7 | 0.070 |
| 10 | PMMA | 0.29 ± 0.04 | $4.47 \cdot 10^{10}$ | 0.00 | 3942 | 0.92 | - | - | 1.1 ± 0.5 | 0.020 |
| 11 | 1%-S | 0.33 ± 0.03 | $1.03 \cdot 10^{13}$ | 0.60 | 499 | 0.66 | 4.9 | 0.65 | 1.1 ± 0.5 | 0.029 |
| 12 | 2%-S | 0.32 ± 0.03 | $4.42 \cdot 10^{13}$ | 0.82 | 391 | 0.51 | 5.7 | 1.06 | 1.2 ± 0.4 | 0.049 |
| 13 | 3%-S | 0.27 ± 0.02 | $1.70 \cdot 10^{13}$ | 0.66 | 482 | 0.60 | 4.7 | 0.76 | 1.2 ± 0.4 | 0.079 |

## 3.2. Uniaxial compression tests

### 3.2.1. Effect of relative density

**Figure 2** shows an example of the nominal stress versus nominal strain curves obtained for the uniaxial compression tests of the solid and cellular nanocomposites. The solid PMMA is compared with the nanocomposite 2%-S, together with their corresponding cellular materials at high relative density (close to 0.5). The solid and cellular materials initially deform in a linear, elastic manner up until the yield point after which softening and subsequent hardening is observed [41]. The elastic (secant) modulus $E$ is measured from the slope of the initial linear region. The compressive yield strength $\sigma_y$ corresponds to the peak load before softening.

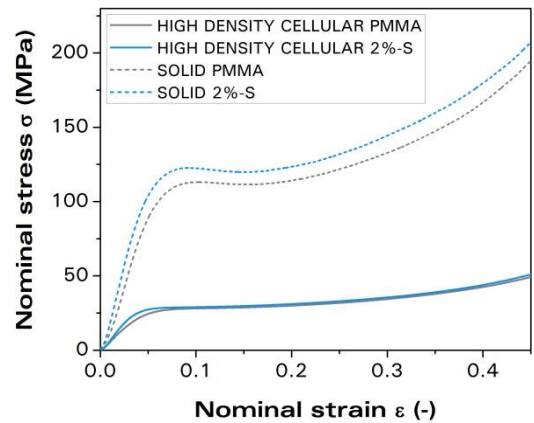

**Figure 2.** Example of stress-strain curves obtained during uniaxial compression of the solid materials and cellular samples with high relative density (around 0.5) based on the PMMA and 2%-S material systems.

Figure 3 shows the elastic modulus and the compressive yield strength of the solid nanocomposites as a function of sepiolite content. It is observed that both properties increase as the sepiolite content increases up to a content in the range of 2 wt% to 3 wt%. Increasing the sepiolite content to 5 wt% does not result in a further increase of the modulus and strength. These trends represent the typical behaviour of polymer nanocomposites: the mechanical properties are enhanced when the filler concentration increases, but there is a critical filler concentration at which there is no further enhancement of the mechanical properties [42]. We observe that the addition of sepiolites induces enhancement of the mechanical properties of the PMMA in uniaxial compression. In particular, for the composite 2%-S, an increase of 15% in the elastic modulus and a 5% in the compressive strength are observed compared to the PMMA without sepiolites. These observations are in agreement with previous reports of an increased strength and modulus when sepiolite particles are added to a polymer matrix [43–45]. In the **Supplementary Information,** several analytical models are used to capture the measured elastic modulus versus relative density trends.

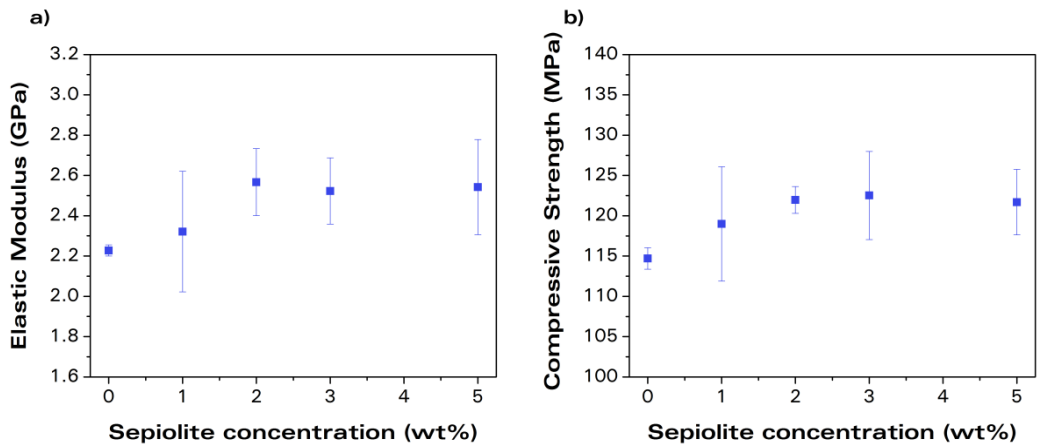

**Figure 3.** a) Elastic modulus and b) compressive yield strength of the solid nanocomposites as a function of sepiolite concentration.

To evaluate the mechanical properties of the cellular materials, the relative elastic modulus ($E_r$) and compressive strength ($\sigma_{y,r}$) are calculated according to equations (5) and (6), respectively, where $E$ and $\sigma_y$ are the properties of the cellular materials and $E_s$ and $\sigma_{y,s}$ are the properties of the solid material with the same sepiolite concentration.

$$E_r = \frac{E}{E_s} \tag{5}$$

$$\sigma_{y,r} = \frac{\sigma_y}{\sigma_{y,s}} \tag{6}$$

The measured trends for the relative modulus versus relative density and the compressive strength versus relative density for the cellular nanocomposites and the cellular PMMA are plotted in **Figure 4.a** and **Figure 4.c**, respectively. Slightly higher values of the relative modulus for the 1%-S, 2%-S, and 3%-S composites at a relative density close to 0.5 are observed, whereas the relative modulus at lower densities is observed to be independent of sepiolite concentration. It was found that the relative yield strength mildly decreases as a function of sepiolite concentration for all investigated relative densities.

It has been reported by several authors [46–49] that a given material property of a cellular polymer ($P_c$) is related to the material property of the solid polymer ($P_s$) by:

$$\frac{P_c}{P_s} = K\rho_r^n \tag{7}$$

Where $K$ and $n$ are constants to be experimentally determined. For most cellular polymers $K$ is close to 1, while $n$ is related to the cellular morphology of the cellular material, being close to 1 for closed cell structures and in the range of 1.5 to 2 for open cell and high density materials [46]. The trends predicted by equation (7) for $K = 1$ are shown in **Figure 4.a** (relative modulus) and in **Figure 4.c** (relative strength) for different values of $n$. One can observe that the relative modulus versus relative density trend of the cellular materials with a high density is captured by equation (7) for $n$ close to 2. In contrast, for the cellular materials with lower relative densities, a $n$ value of 1.5 gives a more accurate fit. The relative compressive strength versus relative density trends (see **Figure 4.c**) are captured by $n$ between 1.5 and 2 for all material systems.

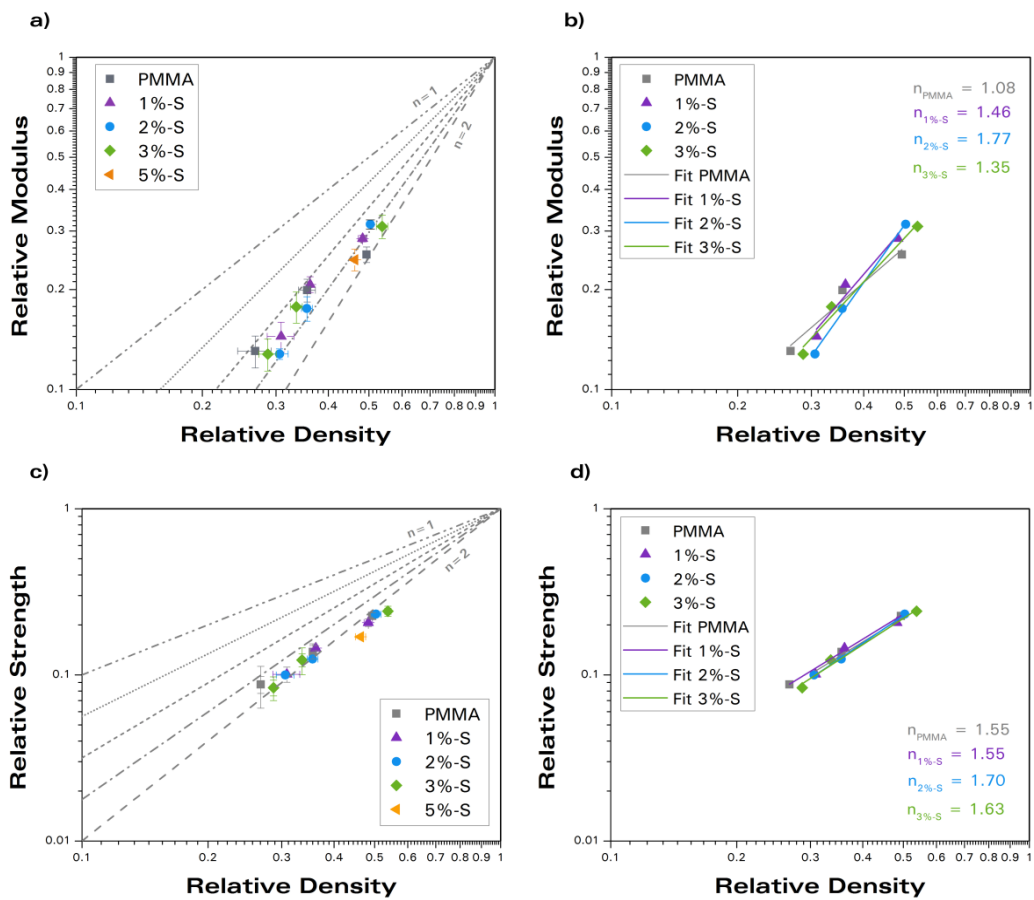

**Figure 4.** a) Relative modulus of the cellular PMMA and the nanocomposites as a function of the relative density with contours predicted by equation (7) for $K = 1$ and $n$ values ranging from 1 to 2; b) Predicted trends by fitting equation (7) to the relative modulus data with corresponding $n$ values; c) Relative compressive strength of the cellular PMMA and the nanocomposites as a function of the relative density with contours predicted by equation (7)

for $K = 1$ and $n$ values ranging from 1 to 2; b) Predicted trends by fitting equation (7) to the relative strength data with resulting fitted $n$ values.

The effect of the relative density is evaluated by fitting equation (7) to the measured relative modulus and relative strength data, giving a fitted $n$ value for each material system with a given sepiolite content (see **Figure 4.b** and **Figure 4.d**). Note that, for this analysis, the system 5%-S was excluded as there were no data points at low densities.

An average $n$ value is calculated from the fitted $n$ values for each material system: $n = 1.42$ for the modulus and $n = 1.61$ for the strength. Equation (7) is then fitted to the measured relative modulus of each material system and the measured relative strength of each material system with the average $n$ by varying $K$. We will use $A$ to denote the $K$ constant for the modulus and $B$ for the $K$ constant for the compressive strength. The obtained values for $A$ and $B$ for each sepiolite concentration are divided by $A_0$ and $B_0$, the value of $A$ and $B$ for the cellular PMMA without sepiolite particles, respectively, as shown in **Figure 5a** (modulus) and **Figure 5b** (strength). The measured modulus of the solid nanocomposite divided by the modulus of the solid PMMA is plotted as a function of the sepiolite concentration in **Figure 5a**. The strength of the solid nanocomposite divided by the strength of the solid PMMA is plotted as a function of the sepiolite concentration in **Figure 5b.** From **Figure 5a** and **Figure 5b** we conclude that, although there is an enhancement of the relative modulus and the relative strength for the solids due to the addition of the sepiolite particles, there is no reinforcement found for the cellular nanocomposites. The trends shown in **Figure 5a** and **Figure 5b** are replotted with error bars in the **Supplementary Information**.

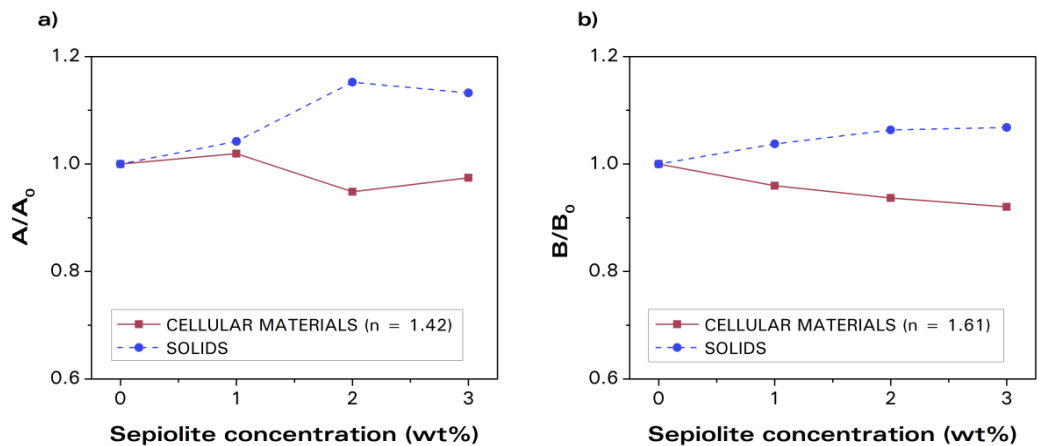

**Figure 5.** a) $A/A_0$ (elastic modulus) and b) $B/B_0$ (compressive strength) as a function of the sepiolite concentration for the cellular materials and the solids.

### 3.2.2. Reinforcement at high relative density

In **Figure 4.a** one can observe that, at high relative densities, the modulus values of the nanocomposites are higher than those of the cellular PMMA. We now perform the same analysis as in Section 3.2.1, but assume $n = 2$. The $n = 2$ assumption for high relative densities (> 0.5) is in agreement with several previous works [48,50–52]. For this analysis, as only the high density materials are considered, the samples with 5%-S are also included. **Figure 6** shows the results of this analysis for the high density materials. One can observe that $A/A_0$ for the cellular nanocomposites with a high density is above unity for all sepiolite concentrations. A clear reinforcement effect is observed for the elastic modulus for the nanocomposites 1%-S and 2%-S, for which the parameter $A/A_0$ takes values as high as 1.18, that is, an 18% enhancement of the modulus by the addition of 2wt% sepiolites. No reinforcement is detected for the compressive strength by assuming $n = 2$ for the high density materials.

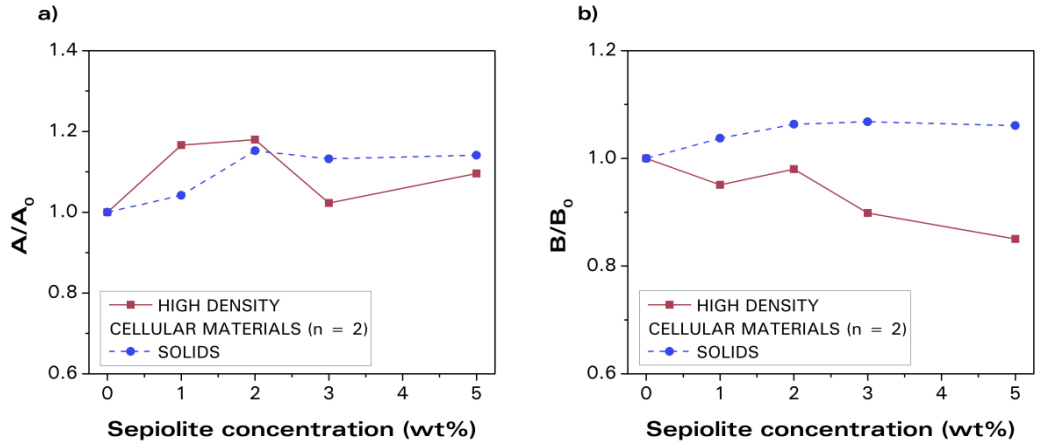

**Figure 6.** a) $A/A_0$ (elastic modulus) and b) $B/B_0$ (compressive strength) as a function of the sepiolite concentration for the cellular materials with high relative density and the solids.

The observed enhancement of the elastic modulus values of the high density materials can be attributed to the presence of the sepiolite particles. Yet, the bimodal cell size distribution and the nano-sized cells could also lead to a potential enhancement of the mechanical properties, see for instance, Notario and colleagues [8] and Miller and coworkers [10]. To verify whether cell size and cell size distribution play a role, additional microcellular materials with 3 wt% of sepiolites were produced and tested in uniaxial compression (see **Supplementary Information**). It was observed that the measured values for the elastic modulus of the bimodal and the microcellular samples were close to each other. These outcomes suggest that the observed enhancement is not caused by the nano-sized cell size and/or the bimodal cell size distribution. Instead, we concluded that the observed reinforcement is due to the addition of sepiolite particles in presence of a cellular structure. This effect was also observed by Laguna-Gutierrez and co-workers who measured the elastic modulus of low density foamed polyethylene reinforced with with silica nanoparticles [53].

Another possible rationale behind the reinforcement detected in the cellular nanocomposites compared to the solid nanocomposites with the same sepiolite content is the improved dispersion of the particles in the cellular materials due to the foaming process. Multiple studies have demonstrated that foaming can lead to better dispersion of particles [54–57]. To validate this hypothesis, the number of particle aggregates was determined before and after the foaming process for the material with 2 wt% of sepiolites (for which the highest enhancement of the modulus was observed at a relative density close to 0.5) using tomography and image analysis. **Figure 7** shows an example of the reconstructed images for the solid and a cellular material with a relative density close to 0.5. The bright dots represent the sepiolites aggregates with dimensions larger than 2.5 μm (corresponding to the spatial resolution of the computed tomography instrument). These aggregates represent 0.57 wt% in the solid material, whereas they only account for 0.15 wt% in the cellular material. Moreover, the number of large aggregates decreases by foaming. These outcomes indicate that the particles are less aggregated in the cellular samples than in the solids. The enhanced dispersion is expected to enhance the mechanical properties of the solid phase. As a result, the reinforcement found for the modulus of the cellular nanocomposites is stronger than in the solid nanocomposites, especially for the systems with 1 wt% and 2 wt% of sepiolites.

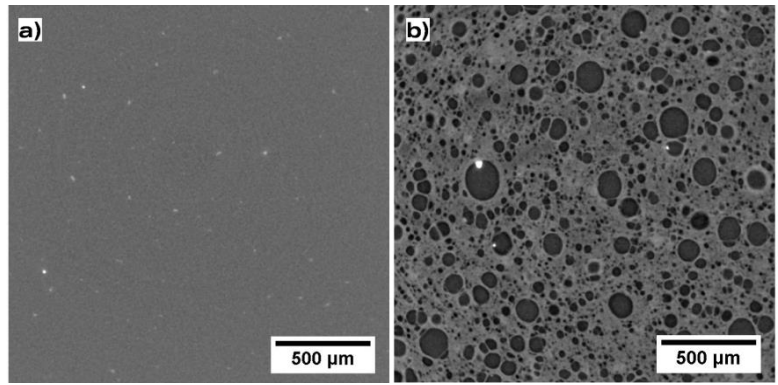

**Figure 7.** Reconstructed tomography images of 2%-S: a) solid nanocomposite and b) cellular nanocomposite with a relative density close to 0.5.

Another potential rationale behind the observed enhancement of the elastic modulus values of the high density materials is related to the position of the aggregates in the cellular materials. Based on SEM micrographs and tomography images, we observe that most of the micro-sized aggregates are isolated from the the solid phase and located within the microcellular pores as a result of the foaming process (see **Figure 8**). The solid phase in the cellular material is therefore reinforced by the small well-dispersed sepiolites, whereas the big aggregates (potentially reducing the mechanical properties of the solid) are not affecting the mechanical performance, as they are located in the microcellular pores. This observation suggests that the presence of a cellular structure in a nanocomposite can balance out, up to some extent, the negative influence of the particle aggregates on the mechanical properties.

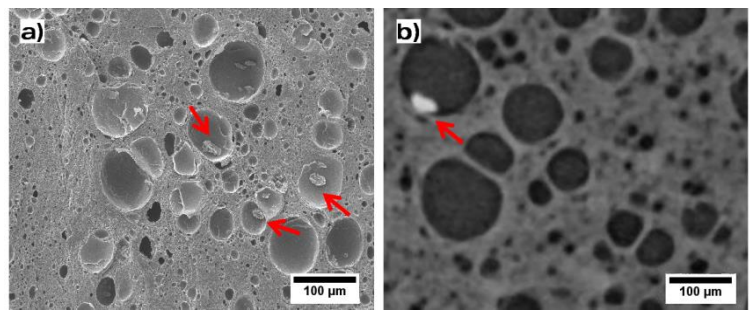

**Figure 8.** Example of aggregates inside the microcellular pores (red arrows): a) SEM micrograph of the cellular material 5%-S with relative density close to 0.5 and b) reconstructed tomography of the cellular material 2%-S with relative density around 0.5.

### 3.3. Fracture Toughness

#### 3.3.1. Effect of relative density on fracture toughness

**Figure 9** shows the measured[1] $K_{IC}$ of the solid nanocomposites and the pure PMMA as a function of sepiolite concentration. The measured fracture toughness of the unfilled PMMA is close to 1.7 MPa m$^{1/2}$, in agreement with reported values for $K_{IC}$ of PMMA in the literature [58]. It is observed that the fracture toughness decreases as the sepiolite content increases. This result is in agreement with earlier works reporting that high aspect ratio fillers such as sepiolites cause embrittlement of the nanocomposite [58,59].

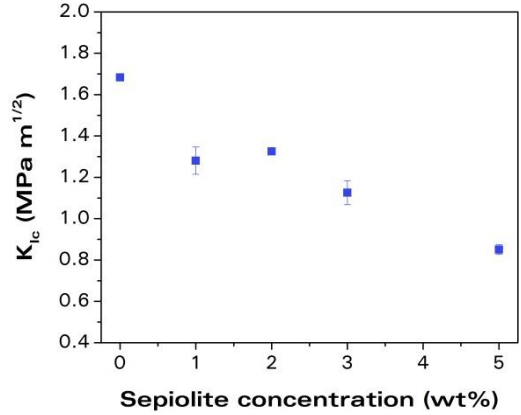

**Figure 9.** Fracture toughness ($K_{IC}$) of the unfilled PMMA and of the solid nanocomposites as a function of sepiolite content.

The fracture toughness of the cellular materials is evaluated by calculating the relative fracture toughness ($K_{IC,r}$) according to equation (8), where $K_{IC}$ is the toughness of the cellular materials and $K_{IC,s}$ is the property of the solid material with the same sepiolite concentration.

$$K_{IC,r} = \frac{K_{IC}}{K_{IC,s}} \quad (8)$$

The trends for the relative fracture toughness versus relative density of the cellular nanocomposites and the cellular PMMA are plotted in **Figure 10.a**. Over the complete density range, the measured relative toughness of the cellular materials with sepiolite particles is higher than the measured relative fracture toughness of the cellular materials without sepiolites. The trends predicted by equation (7) for $K = 1$ are also shown in **Figure 10.a** for different $n$ values.

Equation (7) is fitted to every material system, see **Figure 10.b**. An average $n$ value is calculated from the fitted n values: $n = 1.43$. Equation (8) is subsequently fitted to the measured fracture toughness values for each material system for $n = 1.43$ by varying $K$. We will use $C$ to denote the $K$ constant for the fracture toughness. The obtained $C$ values for each sepiolite concentration are divided by $C_0$, the value for $C$ for the cellular PMMA without sepiolite particles, as shown in **Figure 11.** The fracture toughness of the solid nanocomposites divided by the fracture toughness of the solid PMMA as a function of the sepiolite content is shown in **Figure 11** too. From **Figure 11** we conclude that, although there is a significant decrease of the fracture toughness of the solids as the sepiolite concentration increases, addition of sepiolite particles to the cellular materials leads to an enhanced fracture toughness. This enhancement found in the cellular materials could be a consequence of the presence of a bimodal cell size distribution in combination with nano-sized cell sizes. To evaluate this effect, the fracture toughness of an additional set of microcellular materials with a 3 wt% sepiolite concentration was measured (**Supplementary Information**). It was found that the cell size distribution had no effect on the toughness of the samples. These outcomes therefore suggest that a better dispersion of the sepiolites in the cellular nanocomposites and the presence of the aggregates inside the microcellular pores lead to an improved relative fracture toughness, or in other words, the negative effects of the particle aggregates in the fracture toughness of the solids are hidden in the cellular materials.

---

[1] The load versus indenter displacement trend for all $K_{IC}$ measurements of the solid and cellular material systems was linear up until fracture of the SENB specimens.

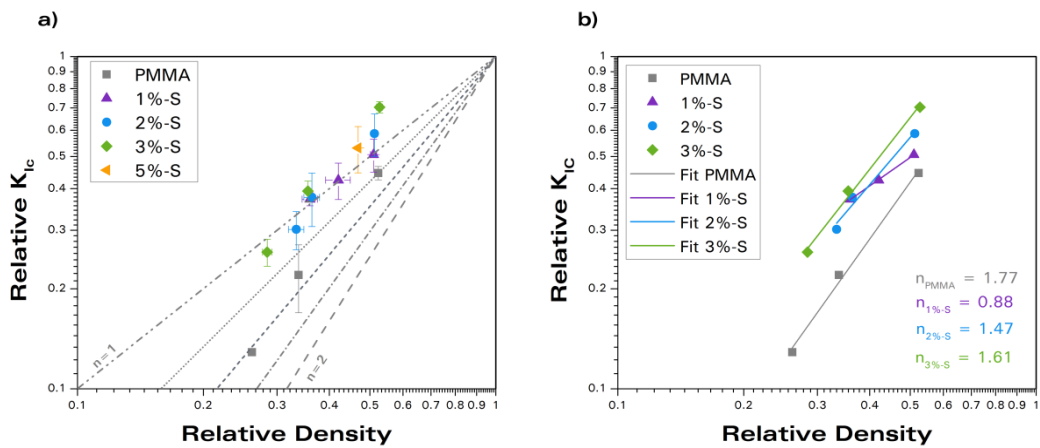

**Figure 10.** a) Relative fracture toughness ($K_{IC}$) of the cellular PMMA and the nanocomposites as a function of the relative density with contours predicted by equation (7) for $K = 1$ and varying *n* values ranging from 1 to 2; b) Fits of the relative modulus according to equation (7) and resulting fitted $n$ values.

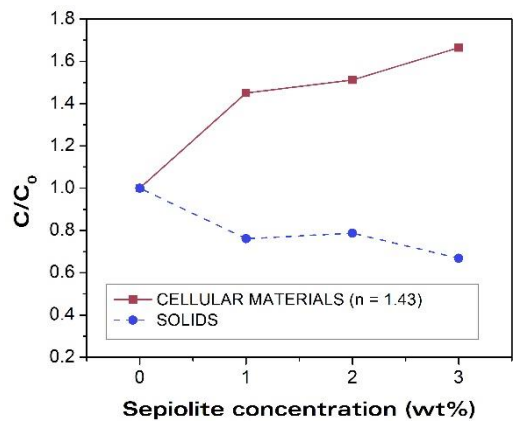

**Figure 11.** $C/C_0$ (fracture toughness constants) as a function of the sepiolite concentration for the cellular materials and the solids.

## 4. CONCLUSIONS

The present study reveals that the addition of up to 3 wt% of nanoparticles made from hydrated magnesium silicates (so-called sepiolites) to solid PMMA leads to a mild increase in modulus (by 15%) and in yield strength (by 5%) but to a decrease in bulk fracture toughness (by 40%). The effect of sepiolite content upon the mechanical properties of PMMA nanocellular materials is more complex. First, the porosity of 50% to 75% exists in a bimodal cell size distribution with one population of cells on the nanoscale and the other on the microscale. The presence of porosity degrades the modulus, strength and toughness for both pure PMMA and for the PMMA-sepiolite composites. In order to isolate the effect of sepiolite content on the relative properties of the foamed PMMA, it is necessary to factor-out the effect of porosity. When this is done, it was found that the relative modulus is independent of sepiolite concentration, whereas the addition of sepiolites results in a mild decrease in relative strength. The relative fracture toughness strongly increases as a function of sepiolite content. Moreover, for the cellular nanocompositites with a relatively low porosity (close to 50%), the addition of sepiolite particles leads to an increase in the relative modulus. Our observations suggest that the enhancement of the relative fracture toughness and the relative modulus (for the porosity of 50%) of the nanocellular PMMA by the addition of sepiolites is caused by the improved dispersion of the sepiolites due to the foaming process and by the migration of the micro-sized sepiolite aggregates to the micro-sized pores during foaming.

**ACKNOWLEDGMENTS**

Financial support from the FPU grant FPU14/02050 (V. Bernardo) from the Spanish Ministry of Education, the Junta of Castile and Leon grant (J. Martín-de León) and the Engineering and Physical Sciences Research Council (UK) award 1611305 (F. Van Loock) is gratefully acknowledged. Financial assistance from MINECO, FEDER, UE (MAT2015-69234-R), the Junta de Castile and Leon (VA275P18), the ERC MULTILAT grant 669764 and SABIC are gratefully acknowledged too. We would also like to thank Dr. Martin van Es from SABIC for the technical assistance and fruitful discussions and Tolsa (Madrid, Spain) for supplying the sepiolites for this study.